# The rights and wrongs of rescaling in population genetics simulations

**Parul Johri[1,2,3,*], Fanny Pouyet[4] and Brian Charlesworth[5]**

[1]Department of Biology, The University of North Carolina at Chapel Hill, Chapel Hill, NC 27599, United States

[2]Department of Genetics, The University of North Carolina at Chapel Hill, Chapel Hill, NC 27599, United States

[3]Integrative Program for Biological and Genome Sciences, The University of North Carolina at Chapel Hill, Chapel Hill, NC 27599, United States

[4]Université Paris-Saclay, CNRS, Laboratoire Interdisciplinaire des Sciences du Numérique, Gif-sur-Yvette, France

[5]Institute of Ecology and Evolution, School of Biological Sciences, University of Edinburgh, Edinburgh EH9 3FL, United Kingdom

*To whom correspondence may be addressed at pjohri@unc.edu

**Abstract**

Computer simulations of complex population genetic models are an essential tool for making sense of the large-scale datasets of multiple genome sequences from a single species that are becoming increasingly available. A widely used approach for reducing computing time is to simulate populations that are much smaller than the natural populations that they are intended to represent, by using parameters such as selection coefficients and mutation rates whose products with the population size correspond to those of the natural populations. This approach has come to be known as rescaling, and is justified by the theory of the genetics of finite populations. Recently, however, there have been criticisms of this practice, which have brought to light situations in which it can lead to erroneous conclusions. This paper reviews the theoretical basis for rescaling, and relates it to current practice in population genetics simulations. It shows that some population genetic statistics are scaleable while others are not. Additionally, it shows that there are likely to be problems with rescaling when simulating large chromosomal regions, due to the non-linear relation between the physical distance between a pair of separate nucleotide sites and the frequency of recombination between them. Other difficulties with rescaling can arise in connection with simulations of selection on complex traits, and with populations that reproduce partly by self-fertilization or asexual reproduction. A number of recommendations are made for good practice in relation to rescaling.

**Author summary**

Large-scale datasets of multiple genome sequences from a single species are increasingly available. Computer simulations of complex population genetic models are essential for making sense of these data, especially for generating a null expectation when testing different evolutionary hypotheses, and for inferring biologically relevant parameters from sequence variation data. Because simulations of large natural populations can be computationally expensive, a widely used tactic that greatly reduces computing time is to simulate populations that are much smaller in size than the populations that they are intended to represent. This approach is known as rescaling, and is justified by population genetics theory, provided some critical assumptions are valid. There has recently been criticism of this practice, highlighting situations in which it can lead to errors. This paper reviews the theoretical basis for rescaling, and relates it to current practice in population genetics simulations. We show that some population genetic statistics are scaleable while others are not. In particular, it shows that there are likely to be problems with rescaling when simulating large chromosomal regions, selection on complex traits, and populations that reproduce partly by self-fertilization or asexual reproduction. A number of recommendations are made for good practice in relation to rescaling.

## Introduction

Computer simulations of complex genetic models that include finite population size effects have a long history in population genetics, dating back to the late 1950s: for some early studies, see Fraser [1]; Fraser and Burnell [2]; Hill and Robertson [3]; Robertson [4]; Franklin and Lewontin [5]. With the advent of large-scale genome sequences from multiple individuals of the same species, computer simulations have become an essential tool for interpreting the data, e.g., [6]. A widely used approach for reducing computing time is to simulate populations that are much smaller than the natural population that they are designed to represent, an approach that has come to be known as rescaling [7,8].

This approach is based on the principle that, provided that the assumptions of diffusion theory (discussed in section 2 below) are met, the evolution of a finite population is determined by the products of the variance effective population size ($N_e$) and the parameters that describe the intensity of the various evolutionary forces (selection coefficients, mutation rates, migration rates, recombination rates), rather than the absolute values of these parameters; for a detailed account of diffusion equation in population genetics, see Ewens [9, Chapters 4 and 5]. This principle was probably first explicitly formulated by Robertson [10] in his seminal paper on the limits to a response to selection, who wrote that "The change in $\phi$ at a particular value of $q$ in an amount of time $t/N$ is dependent only on $Ns$ and on the initial function $\phi(q, 0)$. It then follows that the pattern of the change is determined by $Ns$ and its timescale is directly proportional to $N$." Here, $\phi(q, t)$ is the probability density function for allele frequency $q$ at time $t$, $N$ is the effective population size, and $s$ is the selection coefficient acting on a semi-dominant allele at the locus in question.

The widely used method of Monte Carlo simulations of a diploid Wright-Fisher population of size $N$ assumes random sampling with replacement of gametes from the $2N$ haploid genomes of the surviving adults in a given generation, after the genotype frequencies among the new zygotes have been modified by the deterministic forces under consideration, such as mutation, selection, migration, and recombination. Note that $N$ refers to the number of diploid breeding individuals in a Wright-Fisher population. If $N$ for the simulated population is obtained by dividing the effective size of the corresponding natural population by a factor $C$ ($C > 1$), and the deterministic parameters are multiplied by $C$, the outcome of the simulation with respect to many variables of interest should reflect the behavior of the natural population. As discussed below, such variables are either unaffected by the rescaling or have a known relationship with $C$, in which case the results from the simulation can be adjusted to match the natural population by using the appropriate transformation. The time taken to reach a given state of the population is expected to be divided by $C$ as a result of rescaling, so that two major savings of computer time result from rescaling. The use of the coalescent process for simulating populations and inferring genetic parameters and demographic history similarly assumes that the products of effective population size and deterministic parameters are sufficient to describe the processes involved [9, Chapter 10], [11].

The general validity of rescaling has, however, recently been challenged [12–14], although several studies have shown that rescaling does not affect most summary statistics of interest under neutrality, e.g., [14–16]. Our purpose here is to examine the population genetic justification for rescaling, to determine which evolutionary parameters are likely to be rescaleable and which are not, and to suggest guidelines for deciding on how to use rescaling. Rather than evaluating specific simulation scenarios, we focus on the theoretical foundations of rescaling approaches, with the aim of clarifying their validity for population genetics simulations.

## Results

### The argument from diffusion theory

As the above quotation from Robertson [10] shows, the logic of rescaling comes from the use of diffusion equations to model finite populations [3,10]. These equations are, however, only approximations to the processes that are the basis of most forward simulation methods; as described in the previous section these are usually based on the Wright-Fisher population model with discrete generations, with the new generation produced by random sampling with replacement of gametes, see [9, pp.136-7].

For example, in the case of a single autosomal diallelic locus in a Wright-Fisher population, the column vector $\boldsymbol{f}(t)$ of the probabilities $f_i(t)$ of finding $i$ copies of allele $A_2$ and $2N - i$ copies of allele $A_1$ among a total of the $2N$ alleles present in breeding adults (where $i = 0, 1,\ldots, 2N$) can always be related to $\boldsymbol{f}(t–1)$ by a matrix equation of the form $\boldsymbol{f}(t) = \mathbf{A}\,\boldsymbol{f}(t–1)$, where the elements of $\mathbf{A}$ describe the probabilities of transition between the different frequencies as a result of the action of deterministic forces and the random sampling effects of finite population size [9, Chapter 1]. The diffusion equation approximation passes from a discrete time and discrete frequency representation to a continuous time and continuous frequency one [9, Chapters 1, 4, 5]. This is justified if the changes in the components of $\boldsymbol{f}$ between generations are sufficiently small as to be effectively continuous, and if $2N$ is sufficiently large that allele frequencies are closely packed together on the closed interval [0, 1]. The discrete probability distribution ***f*** for allele frequency $q = i/(2N)$ between 0 and 1 is then replaced by a probability density function $\phi(q, q_0, t)$ (where $q_0$ is the initial value of $q$).

If the expected change in $q$ ($M_{\delta q}$) and the variance in the change in $q$ ($V_{\delta q}$) are of order $1/(2N_e)$, and terms of order $1/(2N_e)^2$ are negligible, the change in $\phi(q, q_0, t)$ per generation is described by the forward Kolmogorov equation [9, Chapters 1, 4, 5]. It is usual to write $V_{\delta q} = pq/(2N_e)$, where $p = 1 - q$ and $N_e$ is the variance effective population size [17]. The following version of the Kolmogorov forward equation then holds:

$$\frac{\partial \phi(q,\ q_0, t)}{\partial t} = \frac{1}{2N_e}\frac{\partial^2 (pq\phi)}{\partial q^2} - \frac{\partial (M_{\delta q}\phi)}{\partial q} \qquad (1)$$

Ewens [9, pp.176-180] discusses the extent to which results from diffusion theory provide accurate approximations to results from the more fundamental Markov chain representation of the Wright-Fisher model described above, including situations when $M_{\delta q}$ is much larger than $1/(2N_e)$. It is important to note that the need for small $M_{\delta q}$ does not necessarily require a variable such as the selection coefficient $s$ in Equation (2) below to be small. For example, if $q$ is confined to values close to zero, as would be the case for a strongly selected deleterious mutation, $M_{\delta q}^2$ would be negligible compared with $M_{\delta q}$ and $V_{\delta q}$. Similarly, in the case of a pair of autosomal loci with recombination frequency $r$ in a randomly mating population, the expected change per generation in the coefficient of linkage disequilibrium due to recombination ($D$) is $M_{\delta D} = -rD$ [18].

In eukaryote reproduction, $r$ normally takes a maximum value of one-half, which applies both to a pair of loci on separate chromosomes and to a pair that are a long way apart on the same chromosome, provided that crossing over occurs at the four-strand stage of the first division of meiosis and that there is no chromatid interference. In theory, chromatid interference could give $r$ values greater than one-half [19] but there is little evidence for its occurrence except in some between-species hybrids in plants [20]. For most biologically realistic cases, a large value of $r$ leads to small values of $D$ unless the population size is very small or selection favoring linkage disequilibrium is strong relative to $r$ [9, Chapter 6]. In this case, it can be assumed that $M_{\delta D}^2$ is negligible, so that diffusion theory can be used even for loosely linked pairs of loci. Similar principles apply to measures of linkage disequilibrium among multiple loci [9, Chapter 6].

Both sides of Equation (1) can be multiplied by $2N_e$, which is equivalent to measuring time in units of $2N_e$ generations, replacing $V_{\delta q}$ with $p(1-p)$, and $M_{\delta q}$ with $2N_e M_{\delta q}$. If $M_{\delta q}$ can be written as the product of a constant and a function of $q$, the constant term is replaced by its product with $2N_e$. For example, the standard model of selection on an autosomal locus in a randomly mating population with selection coefficient $s$ and dominance coefficient $h$ [9, p.13], such that the relative fitnesses of $A_1A_1$, $A_1A_2$ and $A_2A_2$ are 1, 1+ $hs$ and 1+ $s$, gives:

$$M_{\delta q} = spq[h + (1-2h)q] + O(s^2) \qquad (2)$$

Provided that the terms in $s^2$ can be neglected, $s$ can be replaced by $\gamma = 2N_e s$, and the properties of $\partial\phi(q, q_0, t)/\partial t$ are unchanged, except for the change in time-scale, provided that the initial condition $q_0$ is held constant. Note, however, that the dominance coefficient $h$ is not to be rescaled, as it simply modulates the effect of $s$ on $M_{\delta q}$. A similar principle applies to more general genetical situations, such as multiple loci, and to the backward Kolmogorov equation, which is used for solving problems such as the fixation probability of an allele and its expected sojourn time in the population [9, Chapters 4 and 5], as well as the method for calculating changes in the expectations of variables that satisfy the above conditions on $M_{\delta q}$ and $V_{\delta q}$, devised by Ohta and Kimura [21]. Note, however, that fixation probabilities are proportional to $C$ in rescaled populations (Table 1).

If the outcome of the process being studied is independent of the initial conditions, as is the case for the statistics of a stationary probability distribution, such as the moments of the allele frequency $q$ or derived quantities like the genetic diversity as measured by the expectation of $2pq$, the values of these statistics under the diffusion approximation are determined purely by the products of $N_e$ with the parameters describing the changes per generation in allele or haplotype frequencies, provided that these can be written in a similar form to Equation (1).

This principle yields familiar results such as the inverse dependence of the extent of differentiation of neutral allele frequencies between local populations on $4N_e m$ (where $m$ is the migration rate) in Wright's island model [22], or the inverse dependence of the expected magnitude of linkage disequilibrium among a pair of neutral sites on $4N_er$, where $r$ is the recombination frequency [23]. Such quantities would be completely unaffected by rescaling to a different census population size, provided that the assumptions of diffusion theory are met and the products of the deterministic parameters with $N_e$ are held constant. The same applies to properties that depend (to a good level of approximation) on the ratios of deterministic parameters, such as the ratio $u/s$ in deterministic models of mutation selection balance or $r/s$ in models of background selection and selective sweeps [24]. Similarly, properties that are linear functions of $s$ for a given set of genotype frequencies, such as the genetic load [25], are to be scaled by a factor of $C$ in the reduced population, assuming that the genotype frequencies are unaltered by the rescaling, and properties that are quadratic functions of $s$ (such as the genetic variance in fitness) are to be scaled by $C^2$, so that it is easy to transform simulated values to those for the natural population.

There is, however, a problem with properties that depend on the initial conditions, such as the initial frequencies of new mutations. These cannot be held constant if the population size $N$ in a simulation is rescaled by dividing $N$ by a factor of $C >> 1$ and the deterministic parameters such as $s$ are multiplied by $C$, compared with the natural population that is being simulated. It is thus important to examine the conditions under which this problem is likely to affect the outcome of a rescaled simulation. The type of approach involved can be illustrated by the simple case of the rate of substitution, $K$, of new mutations under positive or negative selection in a population with $N$ breeding individuals, assuming that substitutions at different sites in the genome occur independently of each other. The mathematical details are presented in the Appendix (section 1) and summarized in Table 1. Let $N_H$ be the number of haploid genomes among breeding adults ($N_H = 2N$ in the case of an autosomal locus). $K$ is equal to the product of the rate of input of new mutations ($N_H u$) and their probability of fixation, denoted here by $Q(q_0)$, where $q_0 = 1/N_H$ in the case of a new mutation [26, Chapter 1];

The conclusion from the results presented in section 1 of the Appendix is that we can use simulations of a small population to predict the rates of substitution in a large population, provided that the selection coefficient in the reduced population ($Cs$, where $s$ is the selection coefficient for the unscaled population) is sufficiently small that second-order terms in $Cs$ can be neglected. This condition on $Cs$ poses severe limitations on what strength of selection can realistically be modeled; for example, a selection coefficient of – 0.1 for a deleterious mutation in the natural population would turn into a meaningless value of – 10 if $C = 100$. This problem with the magnitude of $s$ will, of course, be encountered in all types of situations involving selection, such as temporally fluctuating selection, background selection, Hill-Robertson interference, and selective sweeps [24].

Other features of interest may, however, be sensitive to initial conditions; for example, section 2 of the Appendix shows that the expected time to loss of a mutation (conditioned on loss) in a population of constant size is a function of $\ln(N_H)$ and hence is not scaleable. In contrast, the expected conditional time to fixation is scaleable. In a finite Wright-Fisher population, the expected number of segregating mutations under the sites model, where at most one variant segregates at a given site in a sequence of nucleotides during its sojourn in the population, is proportional to the product of the number of sites in the sequence and the expected time to loss or fixation of mutations [9, Chapter 9]. Because this time also depends on $\ln(N_H)$, the expected number of segregating sites in the population cannot be rescaled. Furthermore, because the sojourn times of deleterious mutations increase with population size, the effects of background selection on diversity patterns may be affected by rescaling [see 27]. This possibility remains to be further tested.

In contrast, the site frequency spectrum of segregating mutations in a finite sample from the population is scaleable (section 3 of the Appendix), provided that sampling with replacement can be assumed. The same applies to the effects of selective sweeps on neutral diversity under quite general single-locus selection models, provided that $N_e s$ is >> 1; population size affects the results only through $N_e s$ and $N_e r$, at least to a good level of approximation [16,28,29].

**Effects of rescaling on population genetic quantities**

The effects of rescaling on a number of population genetic quantities are summarized in Table 1 and Figure 3. These quantities include parameters associated with mutation rates, selection coefficients and population size, as well as summary statistics measured for the entire population or in a finite sample. The table indicates whether and how these quantities are sensitive to rescaling and provides a framework for interpreting both simulations and theoretical results. For example, rates of processes (e.g. mutation rates) are functions of the scaling factor, $C$, and thus increase with rescaling. However, the same quantities scaled by $2N_e$ are invariant under a change of scale when time is measured in units of $2N_e$ generations.

While the time to fixation (conditional on fixation) of both neutral and selected alleles is preserved with rescaling, the conditional time to loss is not, and increases with the scaling factor (as mentioned above, and shown in Supplementary Figure 1). The dependence on $N_H$ of the fate of mutations implies that summary statistics calculated from the entire population may not be preserved with scaling. In particular, the expected number of segregating variants (Equation A9) and their frequency distribution depends on $N_H$ and is not preserved with scaling.

However, summary statistics for variant frequencies obtained from samples from an equilibrium population are preserved with scaling under the infinite sites model, provided that the sample size is small relative to the population size. Thus, the site frequency spectrum and other statistics like Tajima's *D*, usually should be expected to be preserved with scaling. However, when sample sizes ($n$) are large relative to $N_H$ (i.e., $n^2 \sim N_H$), multiple coalescent events can occur in the external branches of the genealogy, violating the assumptions of the Kingman coalescent and skewing the site frequency spectrum towards more rare variants [30,31].

In general, while summary statistics involving expectations of linear functions of the values of a trait are likely to be preserved with scaling, those involving the variance of a quantitative trait, for instance the variance in fitness between individuals, and genetic and inbreeding load are a function of $C^2$. Thus, they will scale quadratically with the scaling factor, and researchers need to divide them by $C^2$ when comparing simulation results with data. In the following sections, we examine how rescaling behaves in relation to recombination rates, selection on quantitative traits, mating systems, changes in population size and epistasis.

**Effects of rescaling on different simulation methods**

A first type of simulation method, referred to here as the frequency-based method, is to combine the deterministic recursion relations for a given genetic system with a random sampling scheme, either by means of the matrix representation of the probability distribution described above by Equation (1), e.g., [32], or by use of random numbers to simulate sampling from the genotype frequencies generated from the deterministic recursion for a single generation to obtain the transition to the next generation, e.g., [33].

A related procedure is the PoMo (Polymorphism- Aware Phylogenetic Model) method of jointly analysing within- and between-species multiple sequences [34,35]. This uses a single-locus, Moran population genetics model based on the birth-death process [9, pp.104-109] rather than a Wright-Fisher model, with a small, haploid virtual population size that allows rapid computations of the desired statistics, especially phylogenetic tree properties. It can include certain types of selection and biased gene conversion as well as mutation and drift [36]. A transition matrix-based method for handling the properties of single loci that is suitable for analysing the site frequency spectra of very large samples from large populations under the Wright-Fisher model has recently been developed [37].

Another type of method involves forward-in-time, individual-based simulations, including the popular *SLiM* simulation package [38–41]. For example, when simulating a diploid population with no distinction of sex with a Wright-Fisher model of drift, new individuals are generated by randomly sampling pairs of gametes from $N$ adults, following the action of selection, mutation, recombination etc., to produce the next generation of $N$ adults. We refer to this second class as the individual-based method, and use SLiM as an example for most of our discussion.

With all of these methods, other than the PoMo procedure, the size of the population that can be modeled is necessarily limited, either because of the difficulty of computing with large matrices, or because of the time needed to produce $N$ adults, although recent developments have extended computations to larger population sizes and selection coefficients [37].

A third type of method is based on the coalescent process, involving the use of backward-in-time coalescent process simulations [11], which are now being applied to whole chromosomes or genomes by means of various algorithms that represent recombination events [42–44]. These methods are computationally efficient, as only a set of genomes representing a sample of limited size from a large population is involved, and the coalescent process inherently includes the rescaling of the deterministic parameters by $N_e$. However, for large sample sizes or very long genomic regions, the standard coalescent model, which is an approximation to the discrete-time Wright-Fisher (DTWF) model, may create biases in genealogical correlations, identity-by-descent, and long-range linkage disequilibrium. These discrepancies reflect differences between the underlying models rather than incorrect implementation, and recent developments now allow simulation under the exact DTWF model within the *msprime* framework [43].

We therefore focus most of our attention on individual-based simulations. We first consider the most severe problem with rescaling, the rescaling of the rate of recombination.

**The problem with recombination**

The difficulty with rescaling the recombination rate arises from the following considerations. Given that the rate of crossing over between two loci, denoted here by $r$, is restricted by the rules of genetics to $0 \leq r \leq ½$ for most eukaryotes, it would seem to be impossible to multiply $r$ by $C$ for loosely linked or unlinked pairs of loci if we then have $rC > ½$. This is, however, not necessarily a difficulty for the frequency-based simulation methods described above, where the transition between generations for a pair of loci involves $M_{\delta D} = -rD$ , which can be small even if $r$ is arbitrarily large. Since it is $N_e r$ not $r$ that matters for the corresponding diffusion equation process, the biological limitation on $r$ is unimportant, and there is no difficulty in applying rescaling to $r$, provided that only pairwise associations between loci need to be considered. The same applies to methods based on the coalescent process.

Modeling sex differences in recombination should not pose extra problems for rescaling. Basic theory on linkage disequilibrium [9, Chapter 6] suggests that, for autosomes, a simple mean of male and female recombination rates can be used in simulations where the sexes of individuals are ignored; in the extreme case of no recombination in one sex, $r$ calculated from the genetic map in the sex with recombination is simply weighted by ½ for autosomes. For the X or Z chromosomes in species with degenerate Y or W chromosomes, weights of 2/3 and 1/3 for the $r$ values in the homogametic and heterogametic sexes, respectively, should be used [45].

A key consideration when simulating whole genomes or chromosomes is that most finite population models, such as the Discrete-Time Wright–Fisher (DTWF) model, assume no crossover interference [46]. This assumption is made in both individual- and coalescent-based simulations. For instance, in individual-based methods such as *SliM*, random number generators are used to determine the number and location of crossover events on a chromosome. The number of crossovers between a pair of homologous chromosomes is selected from a Poisson distribution, and their locations are assigned by random placement on the chromosome pair. In principle, there is no limit to the number of such crossovers. However, the effect of multiple crossovers is to cause $r$ to increase less than linearly with the map distance $d$ between a pair of loci (in Morgans). For the case of no crossover interference, Haldane's mapping function, $r = \frac{1}{2}[1 - \exp(-2d)]$, applies [47]. It follows that, if we rescale $d$ to $dC$, $r$ is always rescaled by a factor less than $C$, which approaches 1 as $d$ increases. If other parameters, such as selection coefficients and mutation rates, are rescaled by $C$, rescaling introduces a disproportionality between the rate of crossing over and the other deterministic parameters, as well as with drift, compared with the natural population being simulated. Properties such as the effects of hitchhiking that depend on $r/s$, or randomly generated LD for neutral variants (which depends on $N_e r$), or the extent of interference between selected alleles (which depends on $s/r$), will behave incorrectly in the simulations if $d$ not $r$ is rescaled by $C$, as can be seen in the simulations by Marsh et al. [14]. This is bound to affect the outcome of the evolutionary process if multiple loci along a chromosome are being simulated, where it is impossible to rescale every crossover rate by the same $C$. A similar problem applies to recombination between loci on different chromosomes; $r$ for this case is necessarily one-half in a simulation, regardless of $N$, if the process of independent assortment of chromosomes is simulated for each individual in the population.

Complications also arise when using ancestral recombination graph (ARG) methods such as *Relate* [48,49], *tsinfer+tsdate* [50,51], *ARGweaver* [52] or *Singer* [53] to analyse sequence data generated under rescaled forward-time simulations. These methods, which reconstruct local trees under sequentially Markov coalescent (SMC) approximations to the standard coalescent with recombination, rely on the distribution of recombination events along the chromosome. In *ARGweaver* and *Singer*, breakpoints are generated by a Poisson process whose rate depends on the recombination rate per nucleotide site, while in *tsinfer* and *Relate,* breakpoints are approximated via the Li & Stephens [54] method that uses a hidden Markov process representation of recombination.

In simulation-based studies, it is common to infer ARGs from simulated sequence data in order to reproduce the inference pipeline that will later be applied to empirical data, even though the true simulated genealogies are available. In such cases, genealogical summaries are compared between ARGs inferred from simulated data and ARGs inferred from empirical data (non-rescaled). When forward simulations are rescaled, the resulting ARGs may depart from the standard coalescence process, if the rescaled population size becomes too small or if selection is present. This can introduce biases into any analysis that relies on ARG topology, including selection, recombination or demographic inference. In such cases, inferred ARG summaries from rescaled simulations may not be directly comparable to inferred ARG summaries from non-rescaled empirical data. The concern is therefore not that ARG inference methods are intrinsically sensitive to recombination priors, but that strong rescaling may alter the genealogical regime being simulated, potentially affecting the comparability of ARG-based analyses across simulated and real datasets.

There thus seem to be insuperable difficulties in simulating interactions between selection, mutation, recombination and drift that involve whole chromosomes or whole genomes if rescaling is required [14], unless crossing over is rare or absent. For example, in a region under strong background selection, if $N$, $s$, and $r$ are all rescaled proportionally, the ratio $s/r$ at linked neutral sites is expected to be preserved, allowing neutral diversity to reflect the natural population accurately; however, in the presence of multiple crossovers, even proportional rescaling cannot prevent exaggerated reductions in diversity. As shown in Figure 1, the effect of background selection in reducing diversity matches theoretical expectations for small rescaling factors ($C$ = 100 and 200), where only a single and a double crossover is expected per chromosome per generation, but is substantially exaggerated when higher rescaling factors ($C$ = 500 and 1000) result in multiple crossovers (map lengths of 5 and 10 Morgans, respectively). The effects of background selection were larger when the absolute size of the simulated population was smaller, for a given set of scaled parameters and for the same value of $C$ (Supplementary Table 1). This shows that the expectations from diffusion theory fail to apply accurately in this multi-locus context. Even with the larger $N$ values, there was a greater effect of background selection when multiple crossovers were more likely. Overall, for this example the simulated values were fairly close to the theoretically expected $B$ value of 0.74 (based on a linear genetic map) when the simulated population size was not too small, and when the map length in the simulated population was less than two.

The most conservative approach is thus to simulate regions that are sufficiently short that $r$ is nearly linearly related to $d$ after rescaling. This approach is facilitated by crossover interference, whereby the occurrence of a crossover in a bivalent at meiosis I inhibits the occurrence of another crossover nearby [46,55]. Interference reduces the frequency of multiple crossover events, causing the mapping function to become closer to linear. It is feasible to include interference in simulation methods, and this has been done in a study of identity by descent among relatives [56], using an extension of the counting model of Foss et al. [57]. This model assumes that the sites of the initiation of recombination events are distributed randomly along a bivalent at the four-strand stage of meiosis, with a fixed number ($m$) of non-crossover (gene conversion) events separating two successive crossovers, and assuming no interference between sister chromatids. This process generates a gamma distribution of the map distance between successive crossovers in a bivalent, conditioned on a given rate of formation of points of initiation of recombination events, with the scale and shape parameters of the distribution both equal to $m + 1$. Writing $y = 2(m + 1)d$, the mapping function becomes:

$$r = \frac{1}{2}\left[1 - e^{-y}\sum_{i=0}^{m}\frac{y^i}{i!}\left(1 - \frac{i}{m+1}\right)\right] \quad (3)$$

Figure 2 shows the relation between *r* and *d* for several values of *m*, with $m = 0$ corresponding to the Haldane mapping function. It also shows the linear relation between *r* and *d* that results from the case of one crossover per bivalent with complete interference. This mapping function allows an estimate of the maximum length of a chromosome over which an approximately linear relation between *r* and *d* holds after rescaling *d* by a factor of *C*. If the map length of the chromosome in question is *L* and the interference parameter is *m*, the graph of Equation (3) allows visual determination of the value of *d* at which a significant departure from linearity is manifest, denoted by $d_0$. The corresponding proportion of the chromosome is $d_o/L$. For the case of no interference, Figure 2 suggests that $d_o = 0.35$ is a fairly liberal value for this purpose. With $m = 10$, and 4, the corresponding values are 0.45 and 0.40, respectively. The rescaled value of the maximum permitted *d*, $d_1$, must satisfy $d_1 = d_0/C$, and the corresponding proportion of the whole chromosome that can be simulated is thus $d_o/(CL)$. If the physical size of the chromosome is *M* megabases, the corresponding size of a permissible rescaled sequence is $M_C = d_oM/(CL)$.

However, Supplementary Table 1 shows that, as far as the effects of background selection on neutral diversity are concerned, this criterion is probably too stringent – and good results for *B* were obtained even for map lengths of one in small simulated populations. This probably reflects the fact that most effects of selection on linked sites are local, so the total map length of simulated chromosomes is not as restrictive as the linearity requirement implies. Exploratory simulations are thus desirable to determine what combinations of map length and C are likely to yield accurate results. The procedure is simpler in the case of one obligate crossover per bivalent, which is close to what is observed in *C. elegans* [58]. Here, the rescaled maximum permissible map distance is the same as the map length of the whole chromosome (0.5), and the maximum size of a rescaled sequence is simply *M*/*C*.

The extent of interference is highly variable among species [46,59], so that the parameters to be used in a simulation where interference is modeled need to be chosen in relation to what is known about the organism in question. Table 2 provides some examples of species with different levels of interference, with corresponding recommendations for the maximum map lengths consistent with an approximately linear relation between $r$ and $d$, based on the above considerations. These examples are intended only as a very rough guide, as they ignore complexities such as the existence of sex differences in recombination rates, recombination hotspots, regional variation in the rate of crossing over along chromosomes (with greatly reduced rates near telomeres and centromeres being commonly observed), as well as the existence of two pathways to crossing over, only one of which is subject to interference [60].

In *D. melanogaster*, for example, interference is nearly complete over 10cM of the standard female genetic map [61, Chapter 10], which corresponds approximately to four megabases of DNA and contains an average of roughly 500 genes. Without rescaling, a region of this size could be modeled by dropping a single crossover onto a pair of chromosomes with probability 0.1 in females, or a net probability of 0.05 if the absence of crossing over in males is taken into account. However, rescaling of such a region without allowing multiple events would be restricted to a $C$ of 20, giving a probability of one for a crossover event. To rescale by a factor of 1000, which is often used in simulations of Drosophila populations, e.g., [62], the size of the region corresponding to the same probability of a crossover would have to be reduced by a factor of 50, *i.e*, to 80kb, enough to accommodate about eleven typical sized genes. In other words, the larger the rescaling factor, the shorter the size of the region that can be simulated accurately. In contrast, for simulations of human populations, with their much smaller $N_e$, a $C$ value of 10 is quite feasible, so that a 100-fold larger region could be simulated without much loss of accuracy. This advantage is, of course, offset by the low gene density in humans, which means that only 10 typical genes could be accommodated in such a region.

Crossover interference is not currently directly modelled in *SLiM* and other popular population genetics simulation packages. The implementation of crossover interference in simulations is not straightforward [56], especially if one wants to incorporate recombination rate heterogeneity across the genome as well as the existence of two crossover pathways. However, modeling crossover interference allows for more accurate simulations of whole chromosomes, and should be considered in future work.

There will be instances where the phenomenon of interest is localized to a short region such that the effects of the entire genome/chromosome are not necessarily important, so that crossover interference is not relevant. An example is the effects of selective sweeps in a recombining population. Because most effects of sweeps are restricted to small regions, rescaling is unlikely to affect the observed patterns. Marsh et al. [14] found that even large scaling factors in forward-in-time simulations did not result in any deviations in expected summary statistics around the location of the beneficial allele.

Similarly, gene conversion is much less problematical in relation to scaling. It plays an important role in recombination over short distances, as gene conversion tracts in most organisms usually extend over a few hundred basepairs at most, with a mean of 300-459 bp in humans [63,64] and ~440 bp in *D. melanogaster* [65]. It can be modelled by assigning a rate of initiation of a conversion tract, of similar magnitude to the rate of crossing over per basepair, together with an exponentially distributed tract length, such that the rate of recombination between two sites separated by $z$ nucleotides is $2r_g d_g[1 - \exp(-z/d_g)]$, where $r_g$ is the rate of initiation per basepair of gene conversion events and $d_g$ is the mean length of a tract [66]. If the physical organisation of the genome is retained in the simulations, then $d_g$ should be kept constant while $r_g$ is multiplied by $C$.

Many older simulation studies of multi-locus systems assumed that selection coefficients are sufficiently large that the evolutionary processes involved are deterministic, in which case relatively small population sizes can be simulated, e.g., [67]. The level of realism of these studies may, however, be questioned in the light of evidence for small $N_e s$ values for most deleterious mutations from natural populations, e.g., [68,69]. The issue of scaling does not, of course, arise in simulations of natural populations with sufficiently small sizes that running time is not a problem, as in studies of problems in conservation genetics, e.g., [70]. The problems with modeling recombination noted above mean, however, that simulations of polygenic selection (discussed next) must encounter difficulties when large genomic regions are involved.

**Selection on quantitative traits**

Up to now, we have only considered population genetics models that directly assign fitnesses to genotypes. There is, however, considerable interest in the evolutionary dynamics of quantitative traits, where the dynamics of the variants at the individual loci that underly variation in the trait reflect the functional relation between trait and fitness, as well as the relations between the effects of the individual loci and the trait itself [71–74]. Most recent simulation studies of selection on quantitative traits do not use rescaling, e.g., [75,76], with the exception of Hartfield and Glémin [77], whose study system also involves the complication of partial self-fertilization (see below).

Given the current interest in modeling polygenic adaptation [74], the question arise of how to rescale the parameters involved in future simulation studies. We discuss this problem using the familiar nor-optimal or Gaussian fitness model of selection on a single continuously varying trait [74], whose phenotypic value is denoted here by $z$. The following expression represents the fitness of an individual with trait value $z$:

$$w(z) = \exp\left[-\frac{(z-z_0)^2}{2V_s}\right] \qquad (3)$$

where $z_0$ is the optimal value of the trait and $V_s$ is an inverse measure of the strength of selection on the trait, such that smaller values imply a faster decline in fitness as the trait value departs from the optimum.

If each locus has only a small effect on the trait, the rate of change of allele frequency in a given generation in a randomly mating population is determined by the additive and dominance effects of the locus on the trait ($a$ and $d$, respectively), together with the deviation of the trait mean $\bar{z}$ from $z_0$, the phenotypic variance in the trait ($V_z$), and $V_s$ (see Equations A12-A19 of the Appendix). In general, these parameters depend in a complex way on the details of how all the loci concerned determine trait values (dominance, epistasis, allele frequencies, linkage disequilibrium, etc) and will change over time [71,72,73, Chapter 24]. However, in the absence of epistasis and genotype-environment interactions with respect to $z$, the equations should provide a good approximation to what happens over a single generation.

It is shown in the Appendix that, provided that selection is relatively weak (specifically, $V_s >> V_z$), the selection coefficient $s$ at a given locus is inversely proportional to $V_s$, independently of the scale on which $z$ is measured. This result suggests that the factor $C$ by which $s$ should be multiplied when rescaling $N$ by a factor of $C$ can be obtained by dividing $V_s$ by $C$, while leaving the scale of $z$ unchanged. This means that $a$ and $d$ for each locus are unaffected by the rescaling, whereas $s$ is multiplied by $C$. However, this procedure tacitly assumes that the magnitude of $V_s$ has a negligible effect on the properties of the distribution of $z$, which is not exact, due to the effect of selection in generating linkage disequilibrium [71,72,73, Chapter 24], [78].

Similar considerations should apply to the generalization of Equation (3) for multivariate traits under selection [73, Chapter 30], but we have not investigated this question in detail. Given that there are strong conditions on the region of parameter space in which the above conclusion about scaling is valid, and the potential effects of epistasis and genotype-environment interactions have been ignored, rescaling of simulations of selection on quantitative traits should probably be conducted with caution. Further simulation studies that examine these questions are desirable.

**Self-fertilization and facultative sex**

We now consider some problems that arise with mating systems other than conventional random mating. First, the properties of partially self-fertilizing populations have attracted a good deal of attention, due to their importance for the understanding of processes such as the evolution of outcrossing mechanisms in plants [79]. The impact of recombination in populations reproducing by a mixture of selfing and outcrossing is very different from that in randomly mating, outcrossing populations, because (to a good level of approximation), the recursion relation for $D$ is $M_{\delta D} = -(1-F)rD$, where $F$ is the inbreeding coefficient generated by the current frequency of zygotes produced by selfing versus outcrossing [80]. Under strict neutrality, the equilibrium value of $F$ is equal to $S/(2-S)$, where $S$ is the frequency with which zygotes are produced by self-fertilization [81]. It follows that, if $F$ is close to unity, even loci on separate chromosomes have small effective rates of recombination, given approximately by $(1-F)r$. For the first and third types of simulation methods described above, rescaling should present few difficulties, if this transformation is used to represent the frequency of recombination between a pair of loci and only a single pair of loci is modeled.

For individual based simulation methods, however, there are similar problems to those described above (if $S$ is kept constant and $r$ is scaled), unless a heuristic approach of replacing recombination frequencies by $(1-F)r$ is used, in which case $F$ would have to be re-determined from genotype frequencies every generation (note that the rate of selfing, $S$, is not scaleable), since the neutral formula does not necessarily apply when there is selection. Furthermore, $F$ is likely to vary across the genome if there are regional differences in recombination rates, and hence differences in the effects of selection at linked sites. This procedure could thus be problematical, especially as it is unclear whether this heuristic applies to multi-locus linkage disequilibria. Tests of its accuracy based on simulations with different population sizes would be desirable.

Facultative sexual reproduction might be expected to have similar properties to selfing, in that episodes of asexual reproduction are associated with a lack of recombination. Two extreme classes of facultative sex can be envisaged. The first is when there is a constant frequency $\alpha$ of sexual reproduction each generation. The recursion relations for a given genetic system in such a case can be written down as a combination of equations for sexual reproduction (contributing a fraction $\alpha$ of the new zygotes) and asexual reproduction (contributing a fraction $1 - \alpha$), e.g., [82]. This can easily be carried over into any of the simulation methods described above. In the limit of purely asexual reproduction, as in the case of Y or W chromosomes or clonal organisms, the problem of rescaling the rate of recombination does not arise, but it still exists for individual based simulations when $\alpha > 0$, just as in the case of selfing (like $S$, $\alpha$ is not scaleable).

The extreme alternative to this model is when sexual reproduction is episodic, occurring in only a fraction $\alpha$ of generations, and involving every individual in the population. The simplest case is when $\alpha$ is fixed, so that the interval between sexual generations is $1/\alpha$, but variation in $\alpha$ can also be modelled [83]. This mode of reproduction is characteristic of unicellular eukaryotes such as *Chlamydomonas* or *Saccharomyces*, as well as multicellular cyclical parthenogens like *Daphnia.* While simulating such a situation presents no problem in principle, it is unclear how to relate it to a diffusion process, given the abrupt transitions between sexual and asexual reproduction, and hence how to scale $\alpha$.

A recent study of the effect of a single selective sweep on neutral diversity at linked sites in a diploid species found that, provided the mean of $\alpha$ is sufficiently large that several episodes of sex occur during a sweep, the effect of the sweep on nucleotide site diversity is well predicted by the results for a randomly mating population, substituting $r\alpha$ for $r$, where $r$ is the frequency of recombination between a focal neutral site and the target of selection, $\alpha h + (1 - \alpha)$ for $h$, and $\alpha(1 - 2h)$ for $(1 - 2h)$ in Equation (2) [83]. In this situation, $r\alpha$ and $s$ can both be rescaled. However, when at most only one or two episodes of sex occur during the sweep to fixation of an asexual diploid clone that is heterozygous for a beneficial mutation, there is an abrupt phase transition to a situation when diversity relative to the purely neutral case is bounded below by $2r(1 - r)$, with an additional term of approximately $1/(2N_e\alpha)$. The first term represents the fact that a full sweep requires the initial sweep to be followed by a sexual generation, allowing the production of individuals homozygous for the beneficial mutation. The probability that a pair of these inherits non-identical alleles at the neutral site is $2r(1 - r)$, explaining the first term; the second term represents the bulk of the expected increase in diversity during the sweep. In this situation, $r$ is completely unscaleable, whereas $\alpha$ is scaleable. This illustrates how unexpected complexities can emerge with mating systems that differ from the standard of random mating.

**Population size changes**

With nonequilibrium demography modeled by distinct epochs with different population sizes, so that $N_H$ is a specified function of time, denoted by $N_H(t)$, a rescaled value $N_H(t)/C$ can be used in a simulation. To maintain the rate of coalescence at neutral sites for the original population, the duration ($\tau$) of an epoch would also need to be scaled by $C$, so that $\tau = N_H(t)/C$ generations are simulated for the epoch in question. On one hand, such scaling can be highly beneficial as it reduces both the number of breeding individuals and the simulation time, reducing the total computational cost. However, it has certain limitations, especially when the magnitude of change in population size is large and the number of breeding individuals in the population at any time becomes very small. In particular, the assumptions of the diffusion approximation and its corollary, the Kingman coalescent, break down with very small population sizes, with multiple merger coalescents becoming important [84,85]. In the extreme of a strong population bottleneck in a dioecious species, it is not possible to simulate fewer than two breeding individuals, which poses an obvious limit on the size of $C$. Similarly, if the population experiences a size change for a short time period $\tau'$, $C$ must be limited to a value less than $\tau'$ so that at least one generation elapses at the altered population size. It should be possible to scale a simulated population using different scaling factors during different epochs, e.g., one could, in principle, unscale a population during the bottleneck. However, it is unclear whether the dynamics of linkage disequilibrium and multi-locus selection would be accurately represented if this is done. This question requires further investigation. Analytical and computational methods for dealing with the effects of variable population size in single-locus models on the estimation of the distribution of variant frequencies under both selection and neutrality, and applications to demographic inference and estimation of the distribution of mutational effects on fitness, have recently been developed [86].

**Epistasis**

There are two ways in which epistasis enters into population genetic models. The first is epistasis at the level of fitness itself, in which the fitnesses of multilocus genotypes depart from the predictions from additive or multiplicative combinations of the effects of each locus on its own [9, Chapters 6 and 7]. This could pose a problem for rescaling if the fitness differences among genotypes represented in the population after rescaling become so large that the assumption of small changes in allele or haplotype frequencies is violated, or negative fitness values are produced. But this is not substantially different from the problem with large fitness differences encountered in the absence of epistasis.

The other role of epistasis is at the level of the phenotype on which selection acts [87]. The above discussion of selection on quantitative traits suggests that rescaling should be carried out on the measure of the intensity of selection on the trait, without rescaling the trait, but the consequences of epistasis at the trait level for rescaling remain to be explored.

## Discussion

The results described above are intended to shed light on the pros and cons of the use of rescaling in population genetic simulations. The salient conclusions are summarized below and in Figure 4.

1. We strongly recommend that, whenever possible, one should check with available theory whether the population genetic quantity or statistic of interest is preserved with scaling, or if it scales with the scaling factor in a known manner. In both of these cases, one can successfully obtain the unscaled quantity/statistic from rescaled simulations. However, if the quantity of interest is not scaleable, rescaling should be avoided. Table 1 summarizes the scaleability properties of many quantities of interest in population genetics.

2. When simulating multiple linked sites, it is important to decide the length of the region that is likely to preserve the evolutionary dynamics for the problem under investigation, according to the criteria outlined in the section on recombination.

3. We have mostly considered discrete time Wright-Fisher populations with constant population sizes, for which theoretical expectations are well-developed. In situations where these are not available, it is advisable to test a minimal set of scaling factors to ensure that rescaling does not alter key evolutionary dynamics. This check is particularly important for extreme bottlenecks or large magnitudes of changes in population size.

4. For modeling quantitative traits with a nor-optimal selection model, the inverse measure of the strength of selection ($V_s$) should be scaled, not the scale on which the trait is measured. We note that this recommendation is based on a simplified model, which ignores features such as the Bulmer effect, which reduces additive genetic variance due to selection-induced correlations among loci [78]. In practice, the consequences of rescaling should be explicitly investigated, rather than assumed to be fully reliable. The problems with rescaling recombination rates (point 2) also need to be considered in connection with quantitative traits.

5. For simulations of species reproducing by partial self-fertilization, the probability of selfing ($S$) should not be scaled, in contrast to other evolutionary parameters. The effect of the extent of inbreeding on the rate of recombination needs to be considered in deciding on what length of sequence can be modeled in multi-locus simulations (as per our second recommendation).

6. Similar properties apply to species reproducing with constant frequency $\alpha$ of sexual reproduction each generation; the case of diploid populations with episodes of sexual reproduction alternating with asexual reproduction is more complex, and requires special treatment if the frequency of sex is very low (see above).

7. The same applies to simulations involving selection if there is a danger that rescaling will introduce unrealistically low fitnesses of genotypes with large numbers of deleterious mutations. In simulations with selection, it should be possible to compare theoretical expectations of diversity at neutral alleles to simulated observations, and a mismatch is likely to indicate problems with rescaling. However, this approach is only useful when simulating scenarios where theoretical expectations can be calculated. Often highly complex models are investigated by means of forward simulations, where it may be difficult to obtain theoretical expectations. In such cases, it would be best to compare the statistics of interest obtained from simulations with different scaling factors.

## METHODS

Simulations were performed with the forward-in-time simulator SLiM [88]. A discrete generation, diploid Wright-Fisher population of size *N* was modeled. A single 1 Mb region was simulated where half of all new mutations were strictly neutral, while the other half experienced purifying selection with fixed deleterious selective effects ($2N_{scaled}s = 100$) and all mutations were assumed to be semidominant. Selected sites were uniformly distributed across the simulated region. Simulations mimicked a population of *D. melanogaster* where the effective population size ($N_e$) before scaling was assumed to be $10^6$, the mutation rate ($u$) was $3\times10^{-9}$ per site/generation, and the recombination rate ($r$) was $1\times10^{-8}$ per site/generation. A scaling factor ($C$) of 100, 200, 500, and 1000 was applied such that $N_{scaled} = N_e/C$, $u_{scaled}=u\times C$, and $r_{scaled}=r\times C$. Burn-in was simulated for $10N_{scaled}$ generations, and the nucleotide site diversity $\pi$ was obtained for neutral mutations using the whole population, as well as a sample of 100 or 10 genomes for 100 independent replicates.The theoretical value of nucleotide site diversity with background selection relative to that under neutrality ($B$) can be approximated by $\exp(-U/M)$, where $U$ is the mean number of deleterious mutations per diploid genome and $M$ is the map length in Morgans [89,90]. For all our simulations, the theoretically expected values of $B$ was 0.74.

In order to test the effect of sample sizes and the absolute population size on the effects of background selection on neutral diversity, three additional sets of simulations were performed. (i) The natural population size was assumed to be larger, with $N_e = 5\times10^6$, for which only 10 replicates were simulated and all other simulation parameters were equivalent to the ones described above. (ii) The natural population size was assumed to be large ($N_e = 5\times10^6$), while preserving the population-scaled mutation and recombination rates, with $u = 0.2\times3\times10^{-9}$ per site/generation and $r = 0.2\times10^{-8}$ per site/generation. (iii) Population sizes were assumed to be large while preserving the population-scaled mutation and recombination rate (as in set ii). But the length of the simulated region was increased to 5 Mb to match the number of crossovers per individual observed in set i, so that the ratio $U/M$ is preserved.

## DATA AVAILABILITY

All scripts used to replicate the results of the study are available at https://github.com/JohriLab/RescalingPerspective.

**ACKNOWLEDGEMENTS**

We thank Jeffrey Spence and three other anonymous reviewers for their insightful comments. PJ was supported by the National Institute of General Medical Sciences of the National Institutes of Health under award number R35GM154969 issued to PJ. FP was supported by the ANR-25-CE12-4245-01 RECAF awarded to FP. The funders had no role in study design, data collection and analysis, decision to publish, or preparation of the manuscript.

## APPENDIX

### 1. The rate of substitution of mutations

Let $N_H$ be the number of haploid genomes among breeding adults in a discrete generation model ($N_H = 2N$ in the case of an autosomal locus). The rate of substitution of mutations, *K*, is equal to the product of the rate of input of new mutations ($N_H u$) and their probability of fixation, denoted here by $Q(q_0)$, where $q_0 = 1/(N_H)$ for a new mutation represented by a single copy. Diffusion theory provides the following general formula for $Q(q_0)$ [9, p.140]:

$$Q(q_0) = \frac{\int_0^{q_0} \psi(y)\, dy}{\int_0^1 \psi(y)\, dy} \qquad \text{(A1a)}$$

where $\psi(y)$ is independent of $q_0$ and is defined as the following indefinite integral:

$$\psi(y) = \exp - 2 \int \frac{M_{\delta y}}{V_{\delta y}} dy \qquad \text{(A1b)}$$

In general, *K* is *not* given by the product of $N_H u$ and a quantity that is either independent of $N_H$ or proportional to $1/N_H$. If the former property applied, no rescaling would be needed to obtain the value of *K* for the natural population, since the product of rescaled $N_H$ and *u* is independent of population size; in the latter case, the value for the natural population in units of generations could be obtained by dividing the simulation value by *C*. For selection on an autosomal locus, as described by Equation (2), Equations (A1) give the following results:

$$\psi(y) = \exp\{-4N_e s[hy + \tfrac{1}{2}(1 - 2h)y^2]\} \qquad \text{(A2a)}$$

where $s > 0$ for a favorable mutation and $s < 0$ for a deleterious one.

If $h = ½$ (a semi-dominant allele), Equation (A1a) simplifies to the widely-used result of Kimura [91]:

$$Q\left(\frac{1}{N_H}\right) = \frac{[1 - \exp(-\frac{N_e s}{N})]}{[1 - \exp(-2N_e s)]} \approx \frac{\frac{N_e s}{N}}{[1 - \exp(-2N_e s)]} \qquad \text{(A2b)}$$

Provided that $N_e$ and $N$ for the simulated population are divided by the scaling factor $C$ and $s$ is multiplied by $C$, this result implies that $Q$ for the simulated population is greater than that for the natural population by a factor of $C$, provided that $s$ is sufficiently small that the expression for $Q$ is accurate. The rate of substitution for favorable mutations is then approximated by:

$$K = 2Nu\, Q\left(\frac{1}{2N}\right) \approx 2N_e us \qquad \text{(A3)}$$

If $u$ is scaled by multiplication by $C$, $K$ is unaffected by the rescaling, provided that time is measured in units of $2N_e$ generations, or if the rate of substitution obtained from the simulation is divided by $C$ to obtain the natural population value of $K$ in time units of generations. Another way of looking at this result is that $K$ relative to the neutral substitution rate $u$ is independent of the scaling factor; the same applies to the ratio of $Q$ and the neutral fixation probability $1/N_H$.

It is not obvious whether that this type of result holds for all values of $h$, due to the quadratic term in $y$ in Equation (A2a). For a completely recessive, favorable mutation ($h = 0$, $s > 0$), however, the following approximation for $Q$ is valid (Kimura 1964, Equation 10.13):

$$Q\left(\frac{1}{2N}\right) \approx \left(\frac{N_e}{N}\right)\sqrt{\frac{2s}{\pi N_e}} \qquad \text{(A4)}$$

With a scaling factor of $1/C$ for $N_e$ and $C$ for $s$, $Q$ is multiplied by $C$, just as in the semidominant case, so that the same scaling principles apply.

The situation for deleterious recessive mutations or mutations with *h* values other than 0 or ½ can be clarified by using the more general selection equation in which *h* in Equation (2) is replaced by a constant *a* and (1– 2*h*) by a constant *b*. This formulation can be used to describe many different situations, including inbreeding and sex-linkage [92], using $q_0 = 1/N_H$ for a new mutation. We can then write the indefinite integral of $\psi(y)$ in Equation (A1b) as follows (correcting a misprint in Equation A3a of Charlesworth [92]):

$$\int \psi(y)\, dy = y + \sum_{i=1}^{\infty} \frac{(-\gamma)^i}{i!} \int (2ay+by^2)^i\, dy \qquad \text{(A5)}$$

where $\gamma = 2N_e s$.

The definite integral in the numerator of Equation (A1a) can be written as:

$$\int_0^{q_0} \psi(y)\, dy = \frac{1}{N_H} \left\{1 - \frac{\gamma}{N_H}\left[a + \frac{b}{3N_H}\right] + O\left(\left[\frac{\gamma}{N_H}\right]^2\right)\right\} \qquad \text{(A6a)}$$

If $N_H >> 1$ and $N_e$ and $N_H$ are of similar magnitude, we have:

$$\int_0^{q_0} \psi(y)\, dy \approx \frac{1}{N_H}\left(1 - \frac{a\gamma}{N_H}\right) \qquad \text{(A6b)}$$

The term inside the brackets can be neglected if *s* is sufficiently small. The denominator in Equation (A1a) involves the sum of 1 and successive powers of $\gamma$ (see Equation A4b of Charlesworth 2022) and is thus independent of the scaling factor if $N_e s$ is fixed. It thus follows that, to a good approximation, the ratio of the fixation probability to $1/N_H$ is also independent of the scaling factor, just as in the two limiting cases described above. There should thus be no difficulty in using a simulation of a small population to predict the rates of substitution in a large population, provided that the selection coefficient in the small population is sufficiently small that second-order terms in *s* can be neglected. This is consistent with the results of Johri et al. [93] where *Drosophila*-like populations were scaled by *C* = 200.

## 2. Times to loss and fixation

Some other quantities that depend on the initial conditions do not, however, have this desirable property, notably the expected time to loss of a new mutation conditional on its loss measured relative to $2N_e$, denoted by $t^{**}$. To establish this point, it is sufficient to use the following approximation, valid for $\gamma$ of order 1 [92, Equation 3]:

$$t^{**} \approx 2N_H^{-1}\{\ln(N_H) - 1 - \frac{5}{18}\gamma b - \frac{1}{18}\gamma^2[a\left(5a + \frac{7}{2}b\right) + \frac{17}{100}b^2]\} \qquad \text{(A7)}$$

Even in the neutral case, for which $t^{**} = 2N_H^{-1}[\ln(N_H) - 1]$, the dependence on $N_H$ cannot be scaled out, because of the term $\ln(N_H)$. In contrast, the neutral expected conditional time to fixation relative to $2N_e$, is simply $t^* = 2$.

The result for $t^{**}$ implies that the equilibrium expected number of segregating nucleotide sites (denoted by $S$) for a sequence of length $L$ is also non-scaleable, assuming the infinite sites model [94], as can be seen as follows. Under this model, $S$ is assumed to be very small compared with $L$, so that each new mutation can be assumed to arise at a fixed site. $S$ is then given by the product of $2N_e$, the rate of input of mutations $N_H uL$, and the expected sojourn time of a mutation in the population relative to $2N_e$ [9, p.298]. The latter is equal to:

$$t = Q\left(\frac{1}{N_H}\right)t^* + [1 - Q\left(\frac{1}{N_H}\right)]t^{**} \qquad \text{(A8)}$$

In the neutral case, with $Q = 1/N_H$, we obtain the expression:

$$S \approx 4N_e uL\ln(N_H) \qquad \text{(A9a)}$$

A more accurate expression [9, p. 298], which corrects for the inaccuracy involved in equating integration and summation and for slight deviations from the predictions of the diffusion approximation, is:

$$S \approx 4N_e uL[\ln(N_H) + 0.6775] \qquad \text{(A9b)}$$

$S$ is clearly dependent on $N_H$ in a manner that cannot be rescaled. The same applies to the distribution of allele frequencies at segregating sites (the site frequency spectrum or SFS), as can be seen as follows. The expected numbers of sites with derived variants at specified frequencies provides a substitute for a probability distribution [94]. If $t(q, q_0)$ is the density function for the time (in units of $2N_e$ generations) that a mutation spends at frequency $q$ given an initial frequency $q_0$, and $N_H u$ is the rate at which mutations enter the population, the expected number of segregating sites out of $L$ sites that have variants at between frequencies $q$ and $q + \mathrm{d}q$ is $2N_H N_e L u\, t(q, q_0)\, \mathrm{d}q$, with $q_0 = 1/N_H$ [9, pp.198-199].

We can represent the SFS for the population in terms of the fraction of sites at frequency $q$ (where $q = 1/N_H$ to $q = 1 – 1/N_H$ in steps of $1/N_H$) among all segregating sites, noting that the factor $2N_H N_e L u$ cancels from top and bottom:

$$\phi(q, q_0) = \frac{t(q,q_0)}{\sum t(q,q_0)} \qquad \text{(A10a)}$$

In general, therefore, the SFS for the population is dependent on $N_H$. For example, in the neutral case, it is found that $t(q_0) \approx 2N_H^{-1} q^{-1}$ and $\sum t(q, q_0) \approx 2N_H^{-1}[\ln(N_H) + 0.6775]$ [9, p.177] so that (with $q_0 = 1/N_H$) we have:

$$\phi(q, q_0) = \frac{1}{q[\ln(N_H)+0.6775]} \qquad \text{(A10b)}$$

The ratios of $\phi(q, q_0)$ for different $q$ values are, however, independent of $N_H$, as the denominator of this expression is separable into two components, one of which depends only on $N_H$ and the other only on $q$.

**3. Properties of samples from populations**

In contrast to the results for the whole population, the properties of samples from populations at statistical equilibrium are scaleable, given certain assumptions. Under the infinite sites neutral model, the SFS for a sample of $n$ alleles can be found as follows. The probability of observing $i$ copies of the derived variant when $n << N_H$ is proportional to:

$$\frac{n!}{i!(n-i)!}\int_0^1 q^{i-1}\,(1-q)^{n-i} dq \approx \frac{n!}{i!(n-i)!} B(n+1, i) = \frac{1}{i} \qquad \text{(A11a)}$$

where $B(n+1, i)$ is the beta function, $(n-i)!/[(i-1)!n!]$.

To obtain the SFS among segregating sites, we have to normalize by dividing by the sum of the harmonic series $1/i$ from $i = 1$ to $n - 1$, *i.e.*, Watterson's correction factor $a_n$, giving the sample SFS as:

$$P_i = \frac{1}{i a_n} \tag{A11b}$$

In general, if the population SFS is separable into two components, one of which is a function only of $q$ and the other only of $N_H$, as in the neutral case, the same lack of dependence of the sample SFS on $N_H$ should hold. This is always the case for single-locus selection models, as can be seen as follows. Equation (A8) of Charlesworth [92] shows that $t(q, q_0)$ with $q_0 = 1/N_H$ depends on $N_H$ through a factor of $N_H$ that multiplies terms that are a function of $q$ and the scaled selection coefficient $\gamma$. If $t(q, q_0)$ is integrated or summed over the interval $(N_H^{-1}, 1 - N_H^{-1})$, there will be a dependence on $N_H$ for the population SFS, as in Equation (A10a). However, if the SFS is determined from the integral of the product of $t(q, q_0)$ and the binomial sampling formula for $i$ copies of the mutant allele, conditioned on $q$, the normalization of the type used for Equations (A11) means that there is no dependence on $N_H$ in the final expression. Thus, while the population SFS is not scaleable, the sample SFS for derived variants and statistics based on these (e.g. Tajima's $D$) should be scaleable, at least if the infinite sites assumption holds.

**4. Selection on a quantitative trait**

A general expression for selection on a single diallelic locus that makes a small contribution to variation in a quantitative trait controlled by many loci with small effects is given by Bürger [72, p.201], which is based on earlier work of Fisher [95, pp. 104-110], Haldane [96] and Wright [97]. A simplified account is presented here. Consider a single diallelic autosomal locus that affects the trait, and which segregates for alleles $A_1$ and $A_2$ with frequencies $p$ and $q$, respectively. Consider all individuals with expected trait value $z$ if the effects of the locus under consideration are ignored. Let the values of the trait for $A_1A_1$, $A_1A_2$ and $A_2A_2$ be $z+a$, $z+d$ and $z-a$, respectively. If the alleles are semidominant, $d = 0$. Note that these are the trait values averaged over all other genotypes and all environments for the generation in question and are thus not necessarily fixed quantities. Assume that $a$ and $d$ are the same for all values of $z$, which implies an absence of epistatic interactions with respect to other loci affecting the trait, as well as an absence of genotype-environment interactions.

Let $w(z)$ be the mean fitness of individuals with trait value $z$ and $\overline{w}$ be the mean fitness of the population. If $a$ and $d$ are sufficiently small that terms of order $a^3$ and $d^3$ can be neglected, the fitnesses of the three genotypes at the locus can be approximated using Taylor's theorem:

$$w_{11}=\mathrm{E}\{w(z-a)\} \approx \overline{w} - a\,\mathrm{E}\{w'(z) - \frac{a^2}{2}w''(z)\} \qquad \text{(A12a)}$$

$$w_{12}=\mathrm{E}\{w(z+d)\} \approx \overline{w} + d\,\mathrm{E}\{w'(z) + \frac{d^2}{2}w''(z)\} \qquad \text{(A12b)}$$

$$w_{22}=\mathrm{E}\{w(z+a)\} \approx \overline{w} + a\,\mathrm{E}\{w'(z) + \frac{a^2}{2}w''(z)\} \qquad \text{(A12c)}$$

where $w'(z)$ and $w''(z)$ are the first and second derivatives of $w(z)$ with respect to $z$.

Using these expressions, the difference between the marginal fitnesses of $A_2$ and $A_1$, $w_2=pw_{12}+qw_{22}$ and $w_1=pw_{11}+qw_{12}$ (Fisher's "average excess" with respect to fitness), is given by:

$$w_2\text{-}w_1 \approx [a+(p-q)d]w'+\frac{1}{2}(q-p)(a^2-d^2)w'' \qquad \text{(A13a)}$$

where $w'$ and $w''$ are the expectations of $w'(z)$ and $w''(z)$, respectively. The change in the frequency of $A_2$ over one generation is given by:

$$\Delta q = pq(w_2 - w_1)/\overline{w} \qquad \text{(A13b)}$$

so that $s=(w_2\text{-}w_1)/\overline{w}$ measures the strength and direction of selection on $A_2$.

A widely used representation of the relation between phenotype and fitness is to write

$$w(z) = \exp[kf(z)] \qquad \text{(A14)}$$

where $k$ is a measure of the strength of selection and $f(z)$ describes the shape of the relation between fitness and phenotype. The nor-optimal model used below is a special case of this formula. The use of the exponential function avoids the possibility of negative fitnesses. In this case, we have:

$$w' = k\,\mathrm{E}\{f'(z)w(z)\}, \qquad w'' = k\,\mathrm{E}\{[f''(z) + kf'^{2}(z)]w(z)\} \qquad \text{(A15)}$$

It is immediately apparent from these expressions that $w'$ is linearly related to $k$ whereas $w''$ has a quadratic relation, if the expectations are independent of $k$, which may of course not be the case in general. This implies that it could be hard to find a way to rescale $s$ in Equation (A13b) in such a way as to keep $N_e s$ constant, unless selection is sufficiently weak that terms in $k^2$ can be neglected.

This conclusion can be made more explicit by considering the nor-optimal selection model, with optimal trait value $z_0$ and inverse measure of selection strength $V_s \equiv 1/2k$:

$$w(z) = \exp\left[-\frac{(z-z_0)^2}{2V_s}\right] \qquad \text{(A16)}$$

If the trait is normally distributed, this expression yields the well-known expressions for the within-generation changes in mean and variance of $z$, $\Delta\bar{z}$ and $\Delta V_z$ [98]:

$$\Delta\bar{z} = \frac{V_z(z_0 - \bar{z})}{(V_z + V_s)}, \qquad \Delta V_z = -\frac{V_z^2}{(V_z + V_s)} \qquad \text{(A17)}$$

Substituting $w(z)$ from Equation (A16) into Equations (A15), and using Equations (A17), yields the following results:

$$\frac{w'}{\bar{w}} = -\frac{1}{V_s}\mathrm{E}\{(z - z_0)w(z)\} = \frac{1}{V_s}\mathrm{E}\{[z_0(\bar{z} + \Delta\bar{z})]w(z)\} = \frac{z_0 - \bar{z}}{(V_z + V_s)} \qquad \text{(A18a)}$$

$$\frac{w''}{\bar{w}} = \frac{1}{\bar{w}V_s}\mathrm{E}\left\{\left[\frac{(z - z_0)^2}{V_s} - 1\right]w(z)\right\} = \frac{1}{\bar{w}}\mathrm{E}\left\{\left[\frac{(z - z_0)^2}{{V_s}^2}\right]w(z)\right\} - \frac{1}{V_s} \qquad \text{(A18b)}$$

Substitution of these expressions into Equations (A13) shows that, for fixed values of $a$, $d$ and $z_0$-$\bar{z}$, the first term in $s$ is inversely proportional to $V_z + V_s$, whereas the second term has a more complex relation to $V_s$ and $V_z + V_s$. The magnitudes of $V_z$ and $\bar{z}$ reflect the values of the $a$'s and $d$'s at the loci affecting the trait, whereas $V_s$ reflects the relation between $z$ and fitness. $V_s$ can be rescaled by $C$ when population size is rescaled by $1/C$, but this does not produce a proportional change even in the first term in the expression for $s$ unless $V_s >> V_z$, *i.e*, selection is weak, and the model converges on the quadratic deviations model:

$$w(z) \approx 1 - \frac{(z - z_0)^2}{2V_s} \qquad \text{(A19a)}$$

In this case, we have

$$w' \approx \frac{(z_0 - \overline{z})}{V_s}, \qquad w'' \approx -\frac{1}{V_s} \qquad \text{(A.19b)}$$

It follows from these relations and Equations (A12) that $s$ is now proportional to $1/V_s$ for fixed values of $a$, $d$ and $z_0$-$\overline{z}$, so that rescaling should work with this model, or with the nor-optimal model when $V_s \gg V_z$.

**Table 1**: Population genetic quantities of interest for single-locus models and their properties with rescaling. Note that when rescaling is performed as follows we have: $N_{\text{scaled}} = N_H/C$, $u_{\text{scaled}} = u\,C$, $r_{\text{scaled}} = rC$, $m_{\text{scaled}} = mC$, $s_{\text{scaled}}=sC$, and $T_{\text{scaled}}=T/C$, where $N_H$ is the number of haploid genomes among breeding individuals, $u$ is the mutation rate per nucleotide site/generation, $r$ is the rate of recombination per site/generation, and $T$ is an interval of time in generations. Throughout, $N_e$ refers to the effective population size, $L$ to the length of the region, $q$ to the mutant allele frequency, $p$=1-$q$, and $n$ refers to the sample size.

| **Population genetic parameter** | **Theoretical expression** | **Scaleability** |
|---|---|---|
| ***Constants*** | | |
| Dominance coefficient | $h$ | Does not scale (it multiplies the selection coefficient) |
| Selfing probability | $S$ | Does not scale (other deterministic parameters are multiplied by functions of $S$) |
| ***Properties of mutations*** | | |
| Probability of fixation of a neutral mutation | $1/N_H$ | Proportional to $C$ |
| Number of fixations of neutral mutations over time $T$ | $uT$ | Preserved with scaling |
| Expected conditional time to fixation of a neutral mutation | $4N_e$ (see section 2 of Appendix) | Preserved with scaling |
| Expected conditional time to loss of a neutral mutation (relative to $2N_H$ generations) | $2N_H^{-1}[\ln(N_H) - 1]$ | Non-linearly related to $C$ |
| Equilibrium frequency of a non-recessive, strongly deleterious mutation | $\frac{u}{hs}$ | Preserved with scaling |
| Fixation probability ($Q$) of a new semidominant selected mutation | $Q = \frac{N_e s / N_H}{1 - \exp(-2N_e s)}$ | Proportional to $C$ |
| Number of fixations of new semidominant selected mutations over time $T$ | $N_H u \times Q \times T$ | Preserved with scaling |
| Expected conditional time to fixation of a new selected mutation | See Equations (A1a) and (A1b) in Charlesworth [92], where the terms | Preserved with scaling |

| | dependent on the initial frequency cancel out in the final expression for $t^*$ | |
|---|---|---|
| Expected conditional time to loss of a selected mutation (relative to $2N_H$ generations) | Equation (A7) | Non-linearly related to $C$ |
| ***Properties of populations*** | | |
| Variance in fitness between individuals | The variance involves the product of $s^2$ and a function of the genotype frequencies | Proportional to $C^2$ |
| Genetic load | Haldane [99]; Crow [25] | Proportional to $C$ |
| Inbreeding load | Morton et al. [100] | Proportional to $C$ |
| ***Summary statistics for the entire population*** | | |
| Number of segregating sites under neutrality | $4N_e uL[\ln(N_H) + 0.6775]$ | Non-linearly related to $C$ |
| Expected SFS $[\phi_{neu}(q)]$ of neutral mutations at equilibrium | $\frac{2N_e u}{q}$ | Preserved with scaling |
| Expected SFS of neutral mutations at equilibrium conditional on segregation | $\frac{1}{q \ln(N_H)}$ | Non-linearly related to $C$ |
| Expected SFS $[\phi_{sel}(q)]$ of semidominant selected mutations at equilibrium | $\approx \frac{4N_e u[1- e^{-2N_e s(1-q)}]}{q(1-q)[1- e^{-2N_e s}]}$ | Preserved with scaling |
| ***Summary statistics for a population sample*** | | |
| Number of segregating sites under neutrality | $L\int_0^1 (1- q^n - p^n)\phi_{neu}(q)dq$ | Preserved with scaling |
| Expected SFS of neutral mutations at equilibrium | $\frac{1}{i}$<br>$i$ is the allele count in the sample | Preserved with scaling |
| Expected SFS of neutral mutations at equilibrium conditional on segregation | $\frac{1}{i\sum_{j=1}^{n-1}(\frac{1}{j})}$<br>$i$ is the allele count in the sample | Preserved with scaling |
| Expected SFS of semidominant selected mutations at equilibrium | $P_i = \frac{n!}{i!\,(n-i)!}\int_0^1 q^i\, p^{n-i}\, \phi_{sel}(q)dq$ | Preserved with scaling |
| Tajima's $D$ | See section 3 of Appendix | Preserved with scaling |
| $F_{ST}$ | $\frac{Var(q)}{\bar{q}(1-\bar{q})}$ | Preserved with scaling |
| | where $Var(q)$ is the variance in allele frequencies among subpopulations | |

**Table 2**: Interference parameters for several species and corresponding recommended maximum lengths for approximate linearity of the relation between recombination rate and map length, as a function of the scaling factor *C*.

| Organism | Interference Parameter | No. of chrs. | Mean map length | Mean Mb per chr. | Maximum length (Mb) | Source |
|---|---|---|---|---|---|---|
| *C. elegans* | Complete | 6 | 0.5 | 28 | 28/*C* | Hillers et al. [58] |
| Mouse* | $m = 10$ | 20 | 0.73 | 135 | 81/*C* | De Boer et al. [101]; Cox et al. [102] |
| Human* | $m = 4$ | 23 | 1.55 | 139 | 36/C | Broman and Weber [103]; Kong et al. [104] |
| *D. melanogaster*** | $m = 4$ | 5 | 0.55 | 24 | 17/C | Zhao et al. [105]; Ashburner et al. [61] |

* This ignores the evidence for a small contribution from non-interfering crossovers.

** These data are for the arms of the 3 major chromosomes; there is no positive crossover interference across the centromere of the two metacentric autosomes.

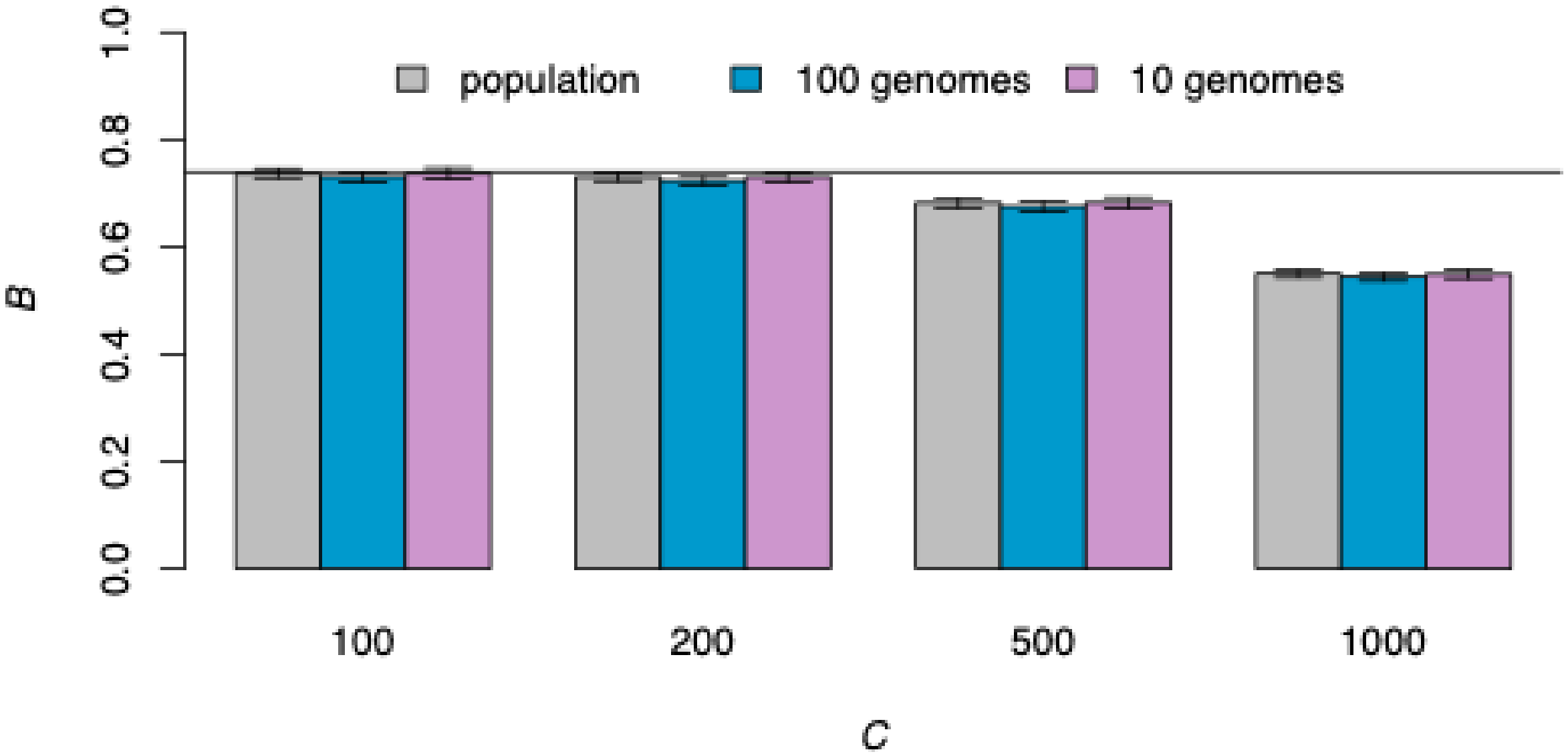

**Figure 1**: *B* is the expected nucleotide diversity at neutral sites in the presence of background selection relative to that under strict neutrality, for different scaling factors (*C*). The solid bars display mean values and the error bars denote the standard errors across 100 independent replicate simulations. Grey bars represent the population means, while blue and purple bars show values when 100 and 10 genomes were sampled, respectively. Forward-in-time simulations were performed for a 1 Mb region where half of all new mutations were neutral, while the other half experienced purifying selection with a constant selective disadvantage ($2N_{scaled}s = -100$) (see Methods). Selected mutations were uniformly distributed across the simulated region. Simulations mimicked a population of *D. melanogaster* where the Wright-Fisher effective population size before scaling was assumed to be $10^6$, the mutation rate was $3\times10^{-9}$ per site/generation, and the recombination rate was $1\times10^{-8}$ per site/generation. The solid black line represents the infinite population value of $B = \exp(-U/M)$ where *U* is the genome-wide diploid mutation rate and *M* is the map length in Morgans, assuming that recombination rate scales linearly with distance. The simulations assumed no crossover interference.

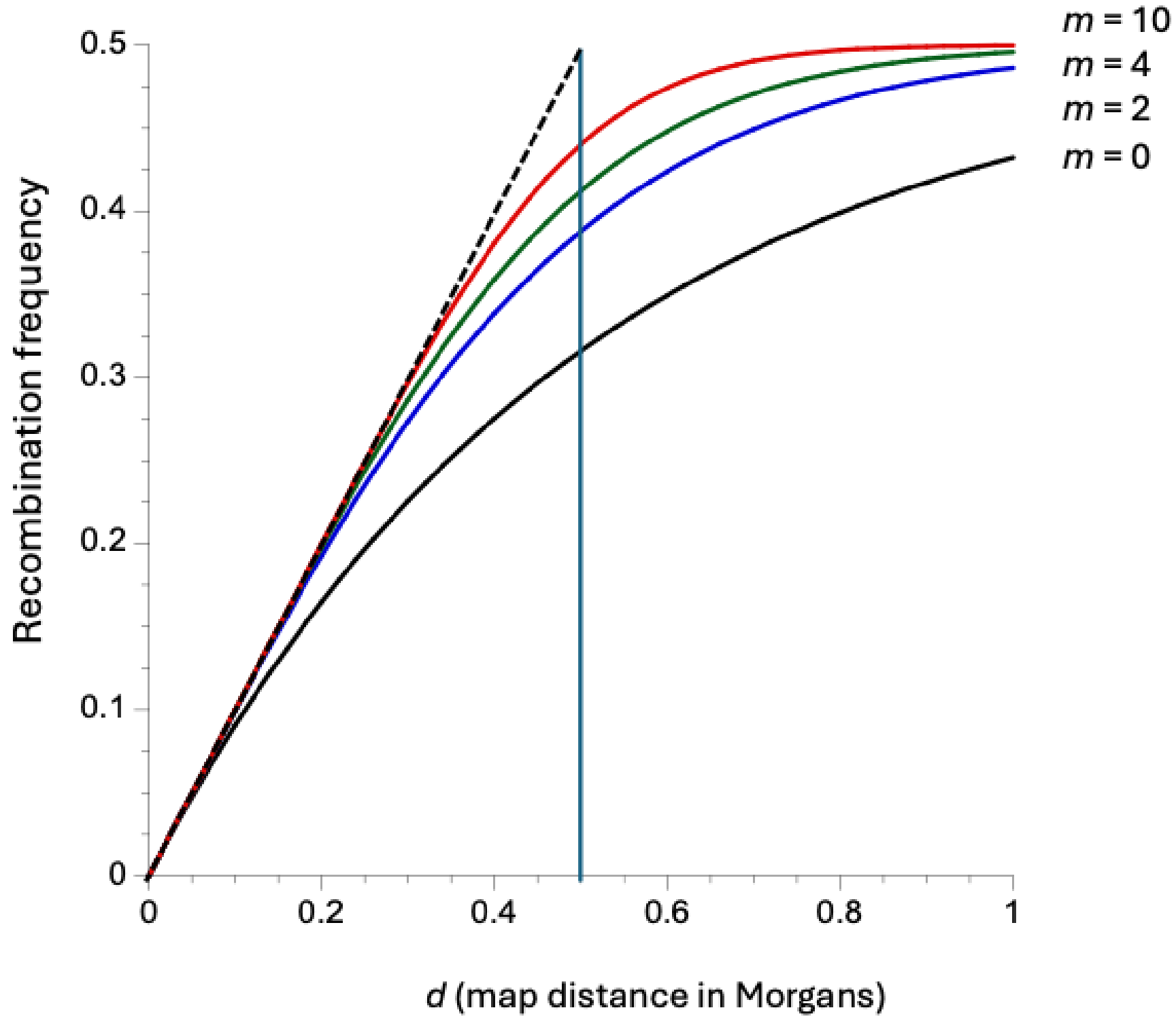

**Figure 2**: Plots of the recombination frequency between two variants as a function of their map distance in Morgans, for four different levels of interference as predicted by the counting model of Foss *et al.* [57]. The parameter $m$ for each curve is the mean number of non-crossover gene conversion events between successive crossovers, with $m$ increasing from bottom to top. The larger $m$, the greater the degree of interference between crossovers; $m = 0$ corresponds to the case of no interference (the Haldane mapping function). The vertical line indicates the map distance of $d = 0.5$ Morgans that corresponds to one crossover per bivalent. The dashed line with a slope of 1 indicated the equality of recombination frequency and map distance when there is complete interference and one crossover per bivalent.

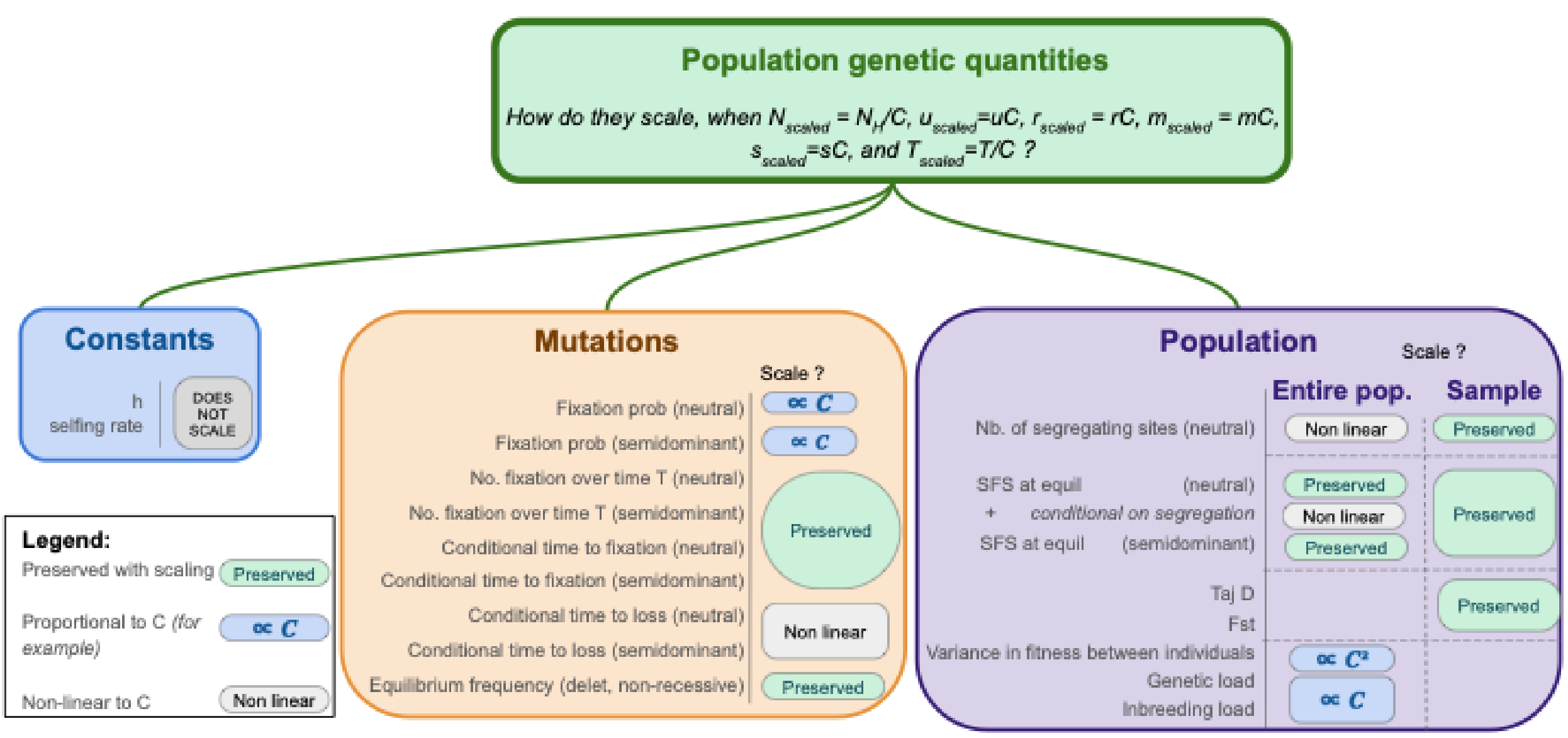

**Figure 3**: A visual summary of Table 1, indicating which summary statistics and quantities are preserved with rescaling.

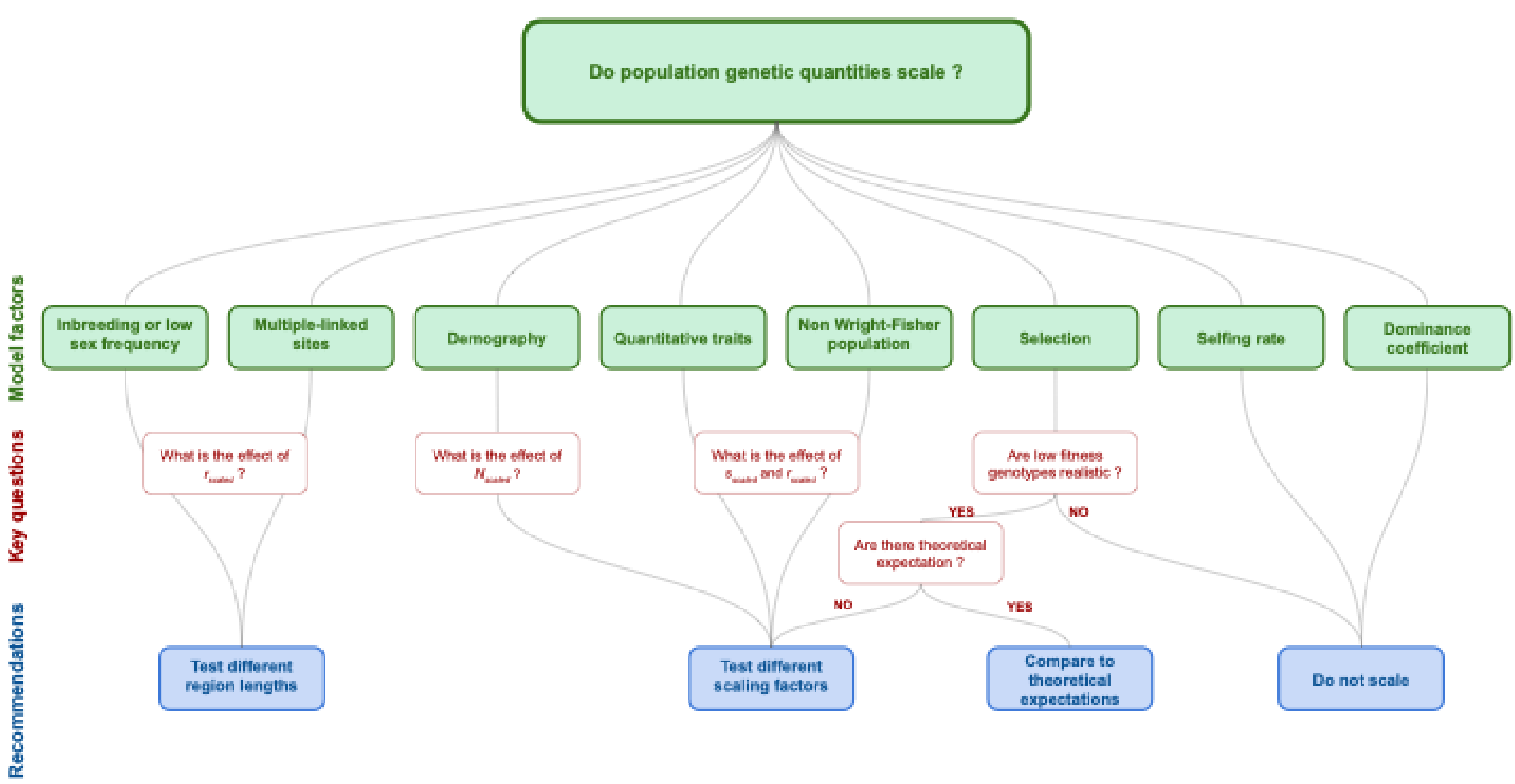

**Figure 4**: A summary of key considerations when using rescaling in simulations.

## SUPPLEMENTARY INFORMATION

**Supplementary Table 1**: Estimated $B$ values in simulations with various scaling factors ($C$) and simulated population sizes ($N_{\text{sim}}$). In all cases, expected $B$ = 0.74. 100 replicates were simulated for simulations with $N$=$10^6$ and 10 replicates were simulated for simulations with $N$=$5\times10^6$ diploid individuals. The assumed mutation rate ($u$) and recombination rate ($r$) are listed per site/generation. The length ($L$) of the simulated region was either 1 Mb or 5 Mb and is specified below. Selected sites experienced purifying selection with a fixed selective disadvantage ($2N_{sim}s$=100). *ML* denotes the map length in the simulated population in Morgans, assuming no crossover interference. SE represents the standard deviation of the means across the simulated replicates. Note that simulations with $N_{\text{sim}}$=25,000 and 50,000 did not reach completion in a reasonable time and thus have been omitted.

**Supplementary Figure 1**: Expected time to conditional loss of neutral mutations (relative to $2N_e$ generations) as a function of the scaling factor plotted using equation A7.

## SUPPLEMENTARY TABLES AND FIGURES

**Supplementary Table 1**: Estimated $B$ values in simulations with various scaling factors ($C$) and simulated population sizes ($N_{sim}$). In all cases, expected $B$ = 0.74. 100 replicates were simulated for simulations with $N$=$10^6$ and 10 replicates were simulated for simulations with $N$=$5\times10^6$ diploid individuals. The assumed mutation rate ($u$) and recombination rate ($r$) are listed per site/generation. The length ($L$) of the simulated region was either 1 Mb or 5 Mb and is specified below. Selected sites experienced purifying selection with a fixed selective disadvantage ($2N_{sim}s$=100). $ML$ denotes the map length in the simulated population in Morgans, assuming no crossover interference. SE represents the standard deviation of the means across the simulated replicates. Note that simulations with $N_{sim}$=25,000 and 50,000 did not reach completion in a reasonable time and thus have been omitted.

| **Parameters** | **Sample size** | **stat** | ***C*=100** | ***C*=200** | ***C*=500** | ***C*=1000** |
|---|---|---|---|---|---|---|
| ***From Figure 1:*** | | | | | | |
| | | | ***ML*=1** | ***ML*=2** | ***ML*=5** | ***ML*=10** |
| | | | **$N_{sim}$=10k** | **$N_{sim}$=5k** | **$N_{sim}$=2k** | **$N_{sim}$=1k** |
| $N_e$=$10^6$<br>$u$=$3\times10^{-9}$<br>$r$=$1\times10^{-8}$<br>$L$=1 Mb | Entire population | mean | 0.737 | 0.730 | 0.682 | 0.550 |
| | | SE | 0.009 | 0.008 | 0.009 | 0.007 |
| | 100 genomes | mean | 0.730 | 0.723 | 0.675 | 0.544 |
| | | SE | 0.009 | 0.008 | 0.009 | 0.007 |
| | 10 genomes | mean | 0.737 | 0.730 | 0.682 | 0.549 |
| | | SE | 0.010 | 0.009 | 0.010 | 0.008 |
| ***Larger natural population size; theta and rho are increased in proportion*** | | | | | | |
| | | | ***ML*=1** | ***ML*=2** | ***ML*=5** | ***ML*=10** |
| | | | **$N_{sim}$=50k** | **$N_{sim}$=25k** | **$N_{sim}$=10k** | **$N_{sim}$=5k** |

| $N_e=5\times10^6$<br>$u=3\times10^{-9}$<br>$r=1\times10^{-8}$<br>$L$=1 Mb | Entire population | mean | - | - | 0.725 | 0.692 |
|---|---|---|---|---|---|---|
| | | SE | - | - | 0.003 | 0.004 |
| | 100 genomes | mean | - | - | 0.718 | 0.685 |
| | | SE | - | - | 0.003 | 0.004 |
| | 10 genomes | mean | - | - | 0.726 | 0.692 |
| | | SE | - | - | 0.003 | 0.005 |
| ***Larger natural population size; theta and rho are preserved*** | | | | | | |
| | | | ***ML*=0.2** | ***ML*=0.4** | ***ML*=1** | ***ML*=2** |
| | | | **$N_{sim}$=50k** | **$N_{sim}$=25k** | **$N_{sim}$=10k** | **$N_{sim}$=5k** |
| $N_e=5\times10^6$<br>$u=0.2\times3\times10^{-9}$<br>$r=0.2\times1\times10^{-8}$<br>$L$=1 Mb | Entire population | Mean | - | - | 0.739 | 0.728 |
| | | SE | - | - | 0.003 | 0.002 |
| | 100 genomes | mean | - | - | 0.731 | 0.720 |
| | | SE | - | - | 0.003 | 0.002 |
| | 10 genomes | mean | - | - | 0.736 | 0.726 |
| | | SE | - | - | 0.003 | 0.002 |
| ***Larger natural population size; theta and rho are preserved; length of the region is increased; the ratio U/M is preserved (M is the map length in Morgans)*** | | | | | | |
| | | | ***ML*=1** | ***ML*=2** | ***ML*=5** | ***ML*=10** |
| | | | **$N_{sim}$=50k** | **$N_{sim}$=25k** | **$N_{sim}$=10k** | **$N_{sim}$=5k** |

| $N_e$=5×10$^6$ $u$=0.2×3×10$^{-9}$ $r$=0.2×1×10$^{-8}$ $L$=5 Mb | Entire population | mean | - | - | 0.726 | 0.691 |
|---|---|---|---|---|---|---|
| | | SE | - | - | 0.001 | 0.001 |
| | 100 genomes | mean | - | - | 0.719 | 0.684 |
| | | SE | - | - | 0.001 | 0.001 |
| | 10 genomes | mean | - | - | 0.726 | 0.690 |
| | | SE | - | - | 0.001 | 0.001 |

**Supplementary Figure 1**: Expected time to conditional loss of neutral mutations (relative to $2N_e$ generations) as a function of the scaling factor plotted using equation A7.

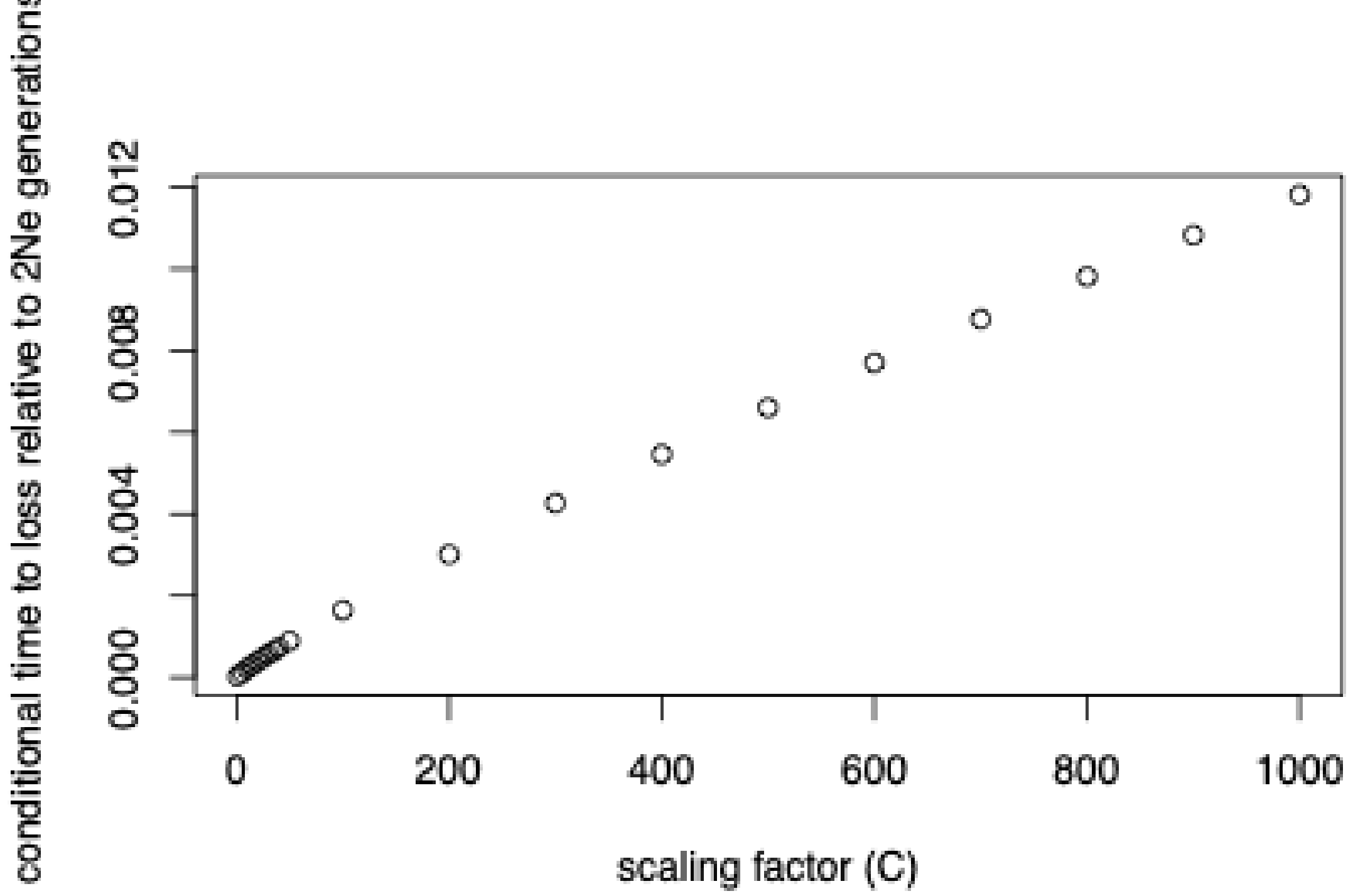